\documentclass[print]{ati}			

\usepackage{graphicx,times}
\usepackage{lmodern}
\usepackage[sort&compress, numbers,super]{natbib} 
\bibpunct[,]{[}{]}{,}{n}{}{,}			
\usepackage{url}
\usepackage[colorlinks=true,breaklinks=true, linkcolor=red,urlcolor=magenta,citecolor=blue,anchorcolor=green]{hyperref}
\usepackage{caption}
\usepackage{rotating}
\usepackage{CJKutf8}	
\usepackage{color}

\renewcommand{\citet}[1]{\citeauthor{#1}(\citeyear{#1})\cite{#1}}	

\def\dB{\rm dB}
\def\MHz{\rm MHz}
   
\begin{document}

   \title{The Electrical
   Design of a Membrane Antenna for Lunar-based Low-frequency Radio Telescope
}

   \author{Suonanben
      \inst{1,2,3,4}
   \and Fengquan Wu\correspondingAuthor{}
     \inst{3,4,5}
    \and Kai He
     \inst{3,4,5}
    \and Shijie Sun
     \inst{3,4,5}
    \and Wei Zhou
     \inst{6,3}
    \and Minquan Zhou
     \inst{7,3} 
   \and Cong Zhang
     \inst{3,4,5}    
     \and Jiaqin Xu
     \inst{3,4,5} 
    \and Qisen Yan
     \inst{7,3}
      \and Shenzhe Xu
     \inst{7,3}
      \and Jiacong Zhu 
      \inst{3,4,5}
       \and Zhao Wang 
      \inst{8}
       \and Ke Zhang
      \inst{8}
      \and Haitao Miao
     \inst{3,4,5}
      \and Jixia Li
     \inst{3,4,5}
      \and Yougang Wang
     \inst{3,4,5}
   \and Tianlu Chen\correspondingAuthor{}
      \inst{1,2}
   \and Xuelei Chen\correspondingAuthor{}
      \inst{3,4,5,9,10}
   }
\correspondent{Fengquan Wu, Tianlu Chen, Xuelei Chen}	
\correspondentEmail{wufq@nao.cas.cn, chentl@utibet.edu.cn, xuelei@cosmology.bao.ac.cn}

\institute{Key Laboratory of Cosmic Rays (Tibet University), Ministry of Education, Lhasa 850000, P. R.China;\email{nbsuo@bao.ac.cn}
          \and
            Department of Physics, College of Science, Tibet University, Lhasa 850000, P. R. China 
          \and
             National Astronomical Observatories, Chinese Academy of Sciences, Beijing 100101, China 
         \and 
             Key Laboratory of Radio Astronomy and Technology, Chinese Academy of Sciences, Beijing 100101, China 
         \and
             School of Astronomy and Space Science, University of Chinese Academy of Sciences, Beijing 100049, China  
        \and
              Shanxi University, Taiyuan 030000, China
         \and
           School of Mechanical Engineering, Hangzhou Dianzi University, Hangzhou 310018, China
         \and
           Center for Astronomy and Space Sciences, China Three Gorges University, Yichang 443002, China
        \and
             Department of Physics, College of Sciences, Northeastern University, Shenyang 110819, China
         \and
             Center of High Energy Physics, Peking University, Beijing 100871, China      
   }
   \date{Received:~February 27, 2024;   Accepted:~April 18, 2024;  Published Online:~May 6, 2024; 
            \DOI{ati2024023} }	
   \copyrights {2024~The Authors. }	
   \citeinfo {Suonanben.et al.2024.}	
   \abstract{
Detecting  primordial  fluctuations  from  the  cosmic  dark  ages  requires  extremely  large  low-frequency  radio telescope  arrays  deployed  on  the  far  side  of  the  Moon.  The  antenna  of  such  an  array  must  be  lightweight,  easily storable  and  transportable,  deployable  on  a  large  scale,  durable,  and  capable  of  good  electrical  performance.  A
membrane  antenna  is  an  excellent  candidate  to  meet  these  criteria.  We  study  the  design  of  a  low-frequency  membrane antenna  for  a  lunar-based  low-frequency  ($<$30  MHz)  radio  telescope  constructed  from  polyimide  film  widely  used  in aerospace  applications,  owing  to  its  excellent  dielectric  properties  and  high  stability  as  a  substrate  material.  We  first design  and  optimize  an  antenna  in  free  space  through  dipole  deformation  and  coupling  principles,  then  simulate  an
antenna  on  the  lunar  surface  with  a  simple  lunar  soil  model,  yielding  an  efficiency  greater  than  90\%  in  the  range  of 12–19  MHz  and  greater  than  10\%  in  the  range  of  5–35  MHz.  The  antenna  inherits  the  omni-directional  radiation pattern  of  a  simple  dipole  antenna  in  the  5–30  MHz  frequency  band,  giving  a  large  field  of  view  and  allowing detection  of  the  21  cm  global  signal  when  used  alone.  A  demonstration  prototype  is  constructed,  and  its  measured
electrical  property  is  found  to  be  consistent  with  simulated  results  using $|S_{11}|$  measurements.  This  membrane  antenna can  potentially  fulfill  the  requirements  of  a  lunar  low-frequency  array,  establishing  a  solid  technical  foundation  for future large-scale arrays for exploring the cosmic dark ages
  }

   \authorrunning{ASTRONOMICAL TECHNIQUES \& INSTRUMENTS }   
   \titlerunning{Suonanben.et al. }  
   \VolumeNumberPageYear{1}{}{227}{2024} 
   \MonthIssue{July}			
   \DOItail{20240215.006}	   	
   \maketitle
   \setcounter{page}{\Page}	
%
%
\section{Introduction}
\label{sect:intro}
The concept of building a radio astronomical observatory  on  the  far  side  of  the  moon  has  been  considered since as early as 1965\cite{Gorgolewski1966}. This is especially vital for the frequency band below 30 MHz, which remains largely unexplored  because  ground-based  observations  are  severely affected by the strong reflection and absorption of Earth’s
ionosphere, together with ubiquitous radio frequency interference  (RFI).  Conducting  low-frequency  radio  observations  on  the  lunar  far  side  or  in  lunar  orbit  not  only avoids  the  influence  of  Earth's  ionosphere  but  also  leverages  the  moon's  shielding  to  block  the  otherwise  substantial  terrestrial  electromagnetic interference\cite{Jester-Falcke2009}. The  opening  up  of  this  new  window  of  astronomical  observation holds  vast  potential  for  new  discoveries  and  promises great  scientific  value.  One  of  the  most  fascinating  potential uses of such a telescope is the observation of the cosmic  dark  ages.  This  is  the  era  after  recombination  of  the
hot plasma following the Big Bang, and before the formation  of  first  generation  stars  and  galaxies\cite{Pritchard-Loeb2012}.The  exploration  and  study  of  the  cosmic  dark  ages  may  give
insights  into  the  early  evolution  of  the  universe  and  the nature of dark matter and offer the opportunity to observe primordial  fluctuations  generated  during  the  inflationary
era  before  their  non-linear  evolution,  elucidating  the  origin  of  the  universe\cite{Loeb-Zaldarriaga2004,Silk2021,Koopmans2021}. Additionally,  the  ultra-long  wave band  can  also  be  used  for  researching  solar  radio  emissions,  exoplanet  radio  emissions,  the  solar  system’s  local environment  in  the  galaxy,  galaxy  medium  distribution,and the origins of cosmic rays, among other subjects \cite{Chen2019}. 

In  1972,  the  US  RAE-2  satellite,  equipped  with  a
low-frequency  inverted-V  antenna,  carried  out  observations  of  ultra-long  wave  signals  in  lunar  orbit,  confirming  extreme  electromagnetic  quietness  on  the  lunar  far
side\cite{Alexander1975}. Subsequently,  a  number  of  spacecraft  have  carried  low-frequency  payloads  to  make  space-based  observations,  primarily  aimed  at  solar  or  planetary  sources(see e.g. \cite{Jester-Falcke2009,Chen2019,Koopmans2021} for further discussions ). During Chinese Chang'e-4 mission, low frequency radio experiments have also been carried out by the lander\cite{Zhu2021}, the relay satellite \cite{ChenL2020}, and the Longjiang-2 lunar micro-satellite \cite{Yan2023}.

Currently,  a  number  of  lunar  radio  astronomy  mission  concepts  are  under  development,  including  lunar orbit  single  satellites  such  as  The  Dark  Ages  Polarimeter Pathfinder (DAPPER)\cite{Burns2021}, or satellite arrays like Discovering  Sky  at  the  ongest Wavelength (DSL)\cite{Chen2021,Huang_2018,Shi_2021,chen_2023}.  It  is  also
a  consideration  in  lunar  surface  project  concepts,  such  as the  Lunar  Surface  Electromagnetics  Experiment (LuSEE)\cite{Bale2023}
,  the  Lunar  Crater  Radio  Telescope
(LCRT)\cite{LCRT2021}, and the Farside Array for Radio Science Investigations of the Dark ages and Exoplanets (FARSIDE)\cite{Burns2021} proposed  by  the  USA,  the  Astronomical  Lunar  Observatory(ALO) project\cite{Koopmans2021}
,  and  the  Large  Array  for  Radio  Astronomy  on  the  Farside  (LARAF)  proposed  by China\cite{Chen2024}.
Projects based in lunar orbit are typically simpler in terms of  space  engineering,  but  the  lunar  surface  provides  certain  advantages;  telescopes  can  be  deployed  on  solid ground,  making  control,  communication,  and  data  analysis  simpler,  especially  if  they  include  a  large  number  of
array elements.


Constructing  a  radio  telescope  with  a  large  receiving area  and  a  large  number  of  elements  on  the  lunar  far  side surface  is  a  great  engineering  challenge  due  to  the  prohibitive expense of shipping the material to the Moon and
the  severely  restricted  amount  of  labor  and  tools  available for deployment and construction. A logical way to alleviate these problems is to construct the antenna using thin films,  which  are  lightweight  and  easy  to  fold  and  store
for  shipping \cite{Jones2013}. Unlike  on  Earth,  where  a  film  needs  to be  fastened  against  the  wind,  lunar  deployment  is  simplified  by  the  lack  of  a  significant  lunar  atmosphere.  This  is
therefore considered as a major design option for most current  lunar  surface  low-frequency  radio  astronomy projects \cite{2003Inflatably,2017Review,Adami_2021}.  Membrane  antennas  have  previously  also been  used  on  Earth  for  other  applications  (e.g.,  cellphone antennas  or  biomedical  antennas),  though  typically  on  a much smaller scale \cite{2011Design}.

In  this  paper,  we  conduct  a  preliminary  study  on  the
electrical  design  and  performance  of  a  membrane  antenna
in  the  context  of  constructing  a  lunar  surface  low-frequency  array.  The  paper  is  organized  as  follows:  in  Section 2, we present the design of the membrane anAtenna; in Section 3, we describe our electromagnetic simulation and
experiment test. We summarize and conclude in Section 4.  

\section{Electrical Design of A Membrane Antenna}
A lunar surface radio telescope can either be a reflector, such as the LCRT\cite{LCRT2021}, an array of dipoles, or any variation  or  combination  of  these.  A  membrane  can  be  applied
in  the  construction  of  any  type.  Here  we  primarily  consider  the  dipole  array  type,  which  is  simpler  to  deploy and  construct  and  is  more  flexible  for  large-scale  applications. The membrane itself can be made with printed conducting layers, which can function as the antenna or connection  wire.  Being  lightweight,  a  relatively  large  geometric
size  can  be  achieved,  at  least  along  one  dimension.Before deployment, the membrane can be rolled to reduce its size for storage. It can then be carried by a lunar rover to the desired site, before being unrolled onto the lunar surface.  A  number  of  materials  can  be  used  for  making  such a  membrane.Polyimide  film,  for  example,  exhibits  excellent  electrical  properties($\epsilon \sim 3.1-3.5, \tan \delta \sim 0.001-0.01$) and has good stability in the ambient temperature  extremes  of  the  lunar  surface\cite{Diaham}.  Consequently,  we selected  it  as  the  antenna  substrate  considered  in  this study.

In the design of membrane antennas, apart from seeking a larger impedance-matched bandwidth, some engineering  and  technical  specifications  must  also  be  satisfied.Here, we assume that for the convenience of rapid deployment  on  the  far  side  of  the  moon,  using  a  reel  on  the lunar  rover,  the  width  of  the  antenna  is  limited,  i.e.,  we assume  the  film  is  a  long  strip.  Wider  films  will  require special  design  to  fold  and  deploy,  which  we  do  not  consider here.  

\begin{figure*}
\centering
\includegraphics[width=\textwidth]{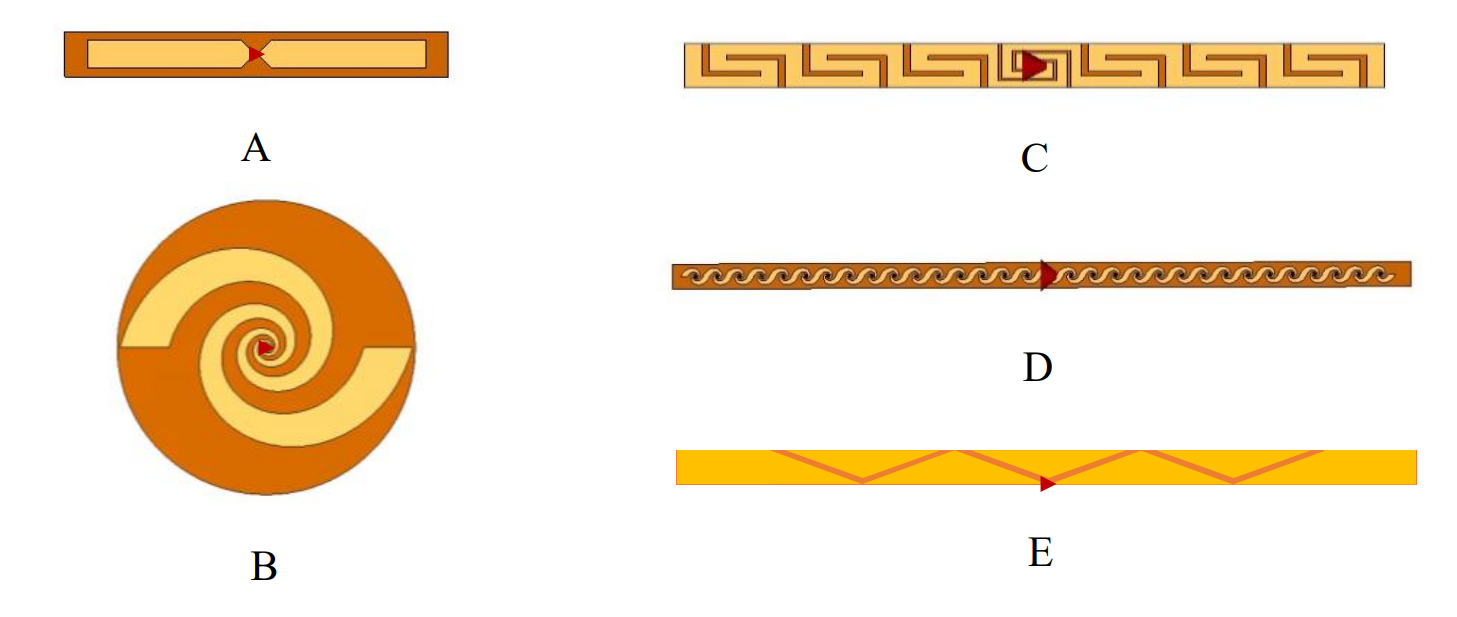}
\caption{Several designs of membrane antennas. A: dipole, B: logarithmic spiral, C: square spiral dipole, D: helical dipole, E: Planar-coupled dipole. The red arrows show the feed port of each antenna}. 
\label{fig:designs}
\end{figure*}

\subsection{Simple Dipole Antennas}

We start with a simple dipole antenna made with the membrane.  as shown in Figure~\ref{fig:designs}-A, with a length of 20 meters, and a thickness of less than 1 millimeter, the  impedance of feed port is 50$~\Omega$. We consider a number of different widths, including 0.01 m, 0.25 m, 0.5 m, 0.75 m, and 1 m. On  each,  we  perform  an  electromagnetic  simulation of  its  electric  performance,  first  in  free  space,  then  with  a lunar  surface  model  (described  in  Section  3.2)  to  investigate its impact on the dipole antenna.

\begin{figure}[h]
    \centering
    \includegraphics[width=0.4\textwidth]{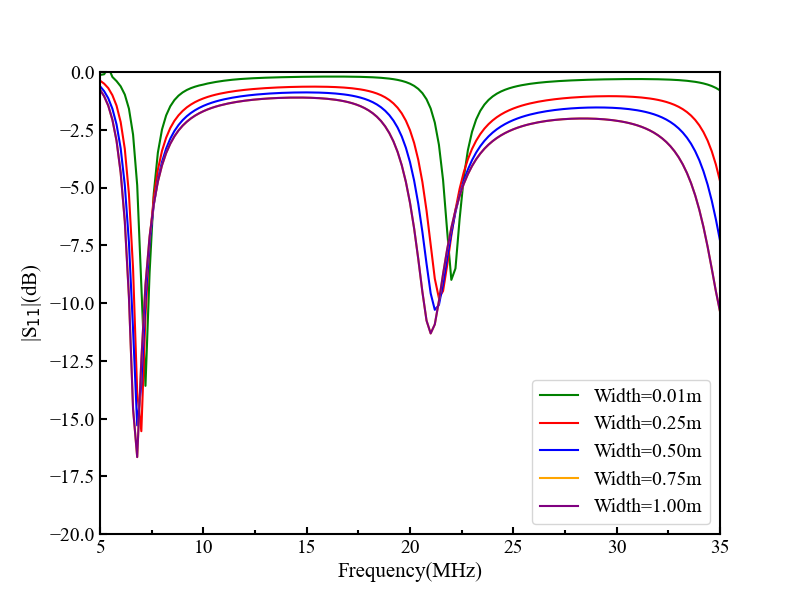}
    \includegraphics[width=0.4\textwidth]{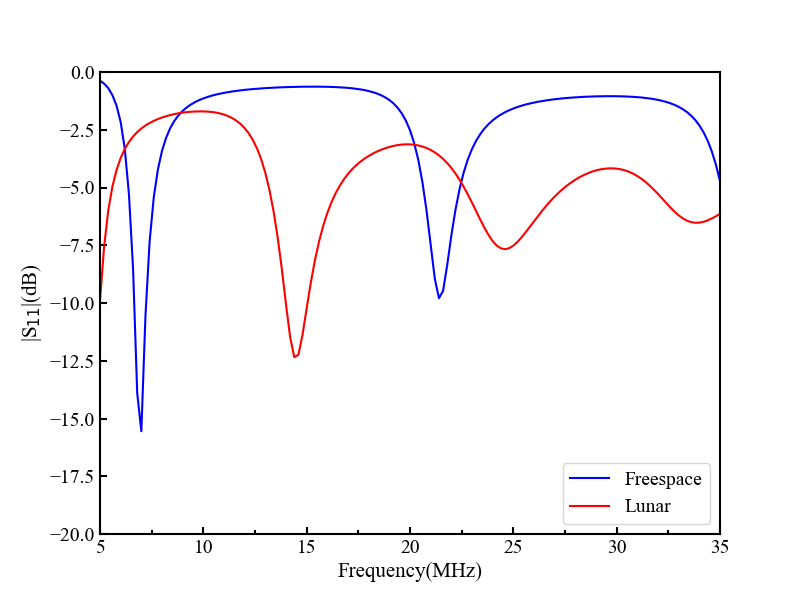}
    \caption{$|S_{11}|$ of the dipole membrane antenna. Top: different width in free space, Bottom: Comparison of the 0.25 m width dipole  in free space and with lunar ground.}
   \label{fig:dipole}
\end{figure}

\begin{figure}[h]
    \centering
    \includegraphics[scale=0.5]{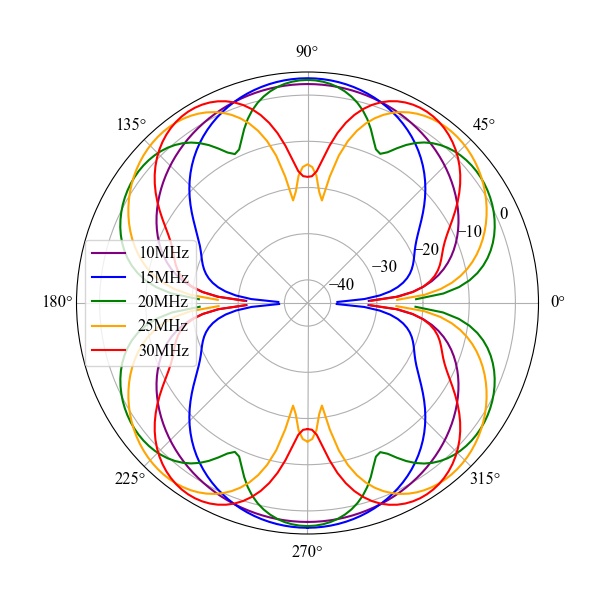}
    \includegraphics[scale=0.5]{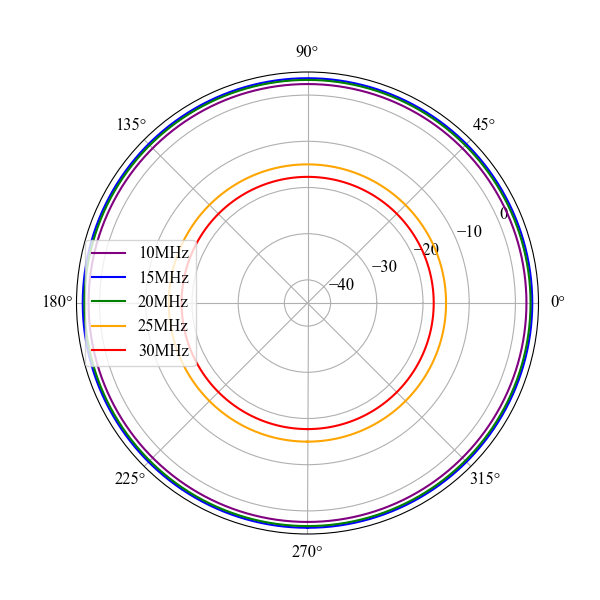}
    \caption{The beam pattern of the dipole model on the E-plane (top) and H-plane (bottom),simulated in free space. } 
    \label{fig:freebeampattern}
\end{figure}

\begin{figure}[h]
    \centering
    \includegraphics[scale=0.5]{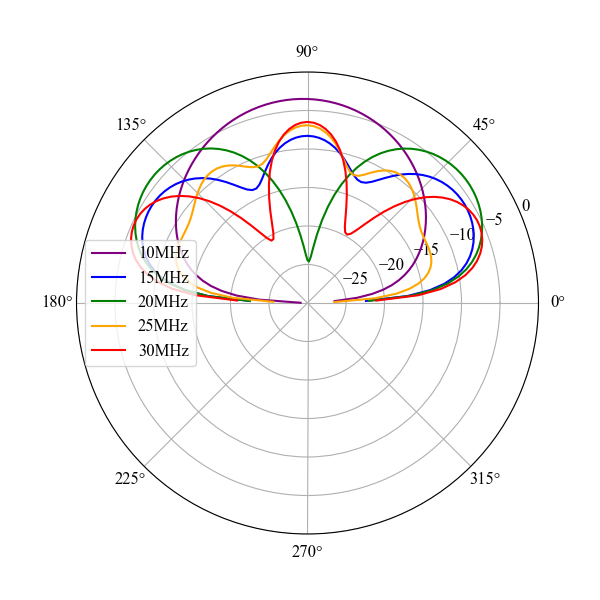}
    \includegraphics[scale=0.5]{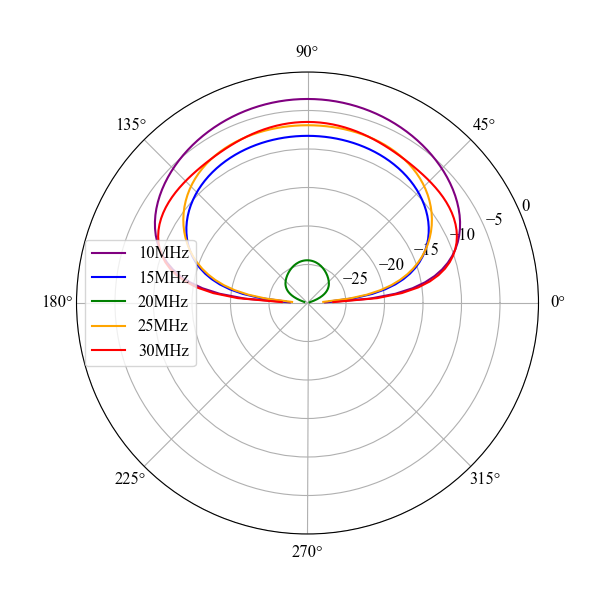}
    \caption{The beam patterns of the dipole model on the E-plane (top) and H-plane (bottom), simulated on a modeled lunar surface. } 
    \label{fig:lunarbeampattern}
\end{figure}

The simulation results of $|S_{11}|$ parameter, as shown in Figure \ref{fig:dipole}, demonstrate that the simple dipoles exhibits narrow resonances in free space. Fig. 2A compares dipoles with different  widths  in  free  space.The  thinnest  has  the  narrowest  resonances,  with  the  fundamental  frequency  occurring where the total length is equal to $\lambda/2=$total length (for $\lambda/2= 20$m, the corresponding frequency is 7.5 MHz), Higher  resonances are  observed  at  odd  multiples  of  the  fundamental  frequency, with the next falling at $3\times 7.5 \MHz = 22.5 \MHz$. For  dipoles  with  larger  widths,  resonance  frequencies  are lower,  with  broader  widths.  We  select  the  0.25  m  width model  as  our  reference  model,  with  its  resonance  occurring at approximately 7 MHz. 

we observed that under the influence of the lunar surface (bottom panel of Fig.~\ref{fig:dipole}), the resonance points of the dipole antenna shift to lower frequencies compared to free space. This is because the lunar surface environment increases the effective permittivity of the environment in which the dipole is located, compared with free space.

Simulated antenna beam patterns are shown in figure~\ref{fig:freebeampattern} for free space , and Figure~\ref{fig:lunarbeampattern} for the lunar surface. In  free space,  the  antenna  maintains  a  broad  dipole  pattern.  As the  frequency  increases,  the  two-lobed  beam  pattern  splits into four lobes, due to the reverse currents on the antenna at  shorter  wavelengths.  The  presence  of  the  lunar  surface causes  the  antenna  pattern  under  the  lunar  surface  almost disappear.  Owing  to  the  effect  of  surface  loss,  the  peak gain  of  the  dipole  antenna  on  the  lunar  surface  is  usually below  0  dBi.  At  higher  frequencies,  the  antenna  pattern will split further, giving more lobes.

\subsection{Design Schemes of Membrane Antennas}
\label{sect:data}
A  simple  dipole  has  a  narrow  working  band  which requires  a  small $|S_{11}|$,  making  it  not  very  suitable  for  the
purpose  of  observing  low  frequencies  in  a  range  of  astronomical  applications.  This  is  a  general  problem  in  the
design of low-frequency receivers. These have a large relative  variation  of  frequencies,  making  it  difficult  to  maintain  well-matched  impedance  of  the  antenna  and  receiver
over a wide frequency range. Here, we try to design a system  to  achieve  good  impedance  matching  over  a  relatively wide bandwidth.
 



\begin{figure*}
    \includegraphics[width=0.9\textwidth]{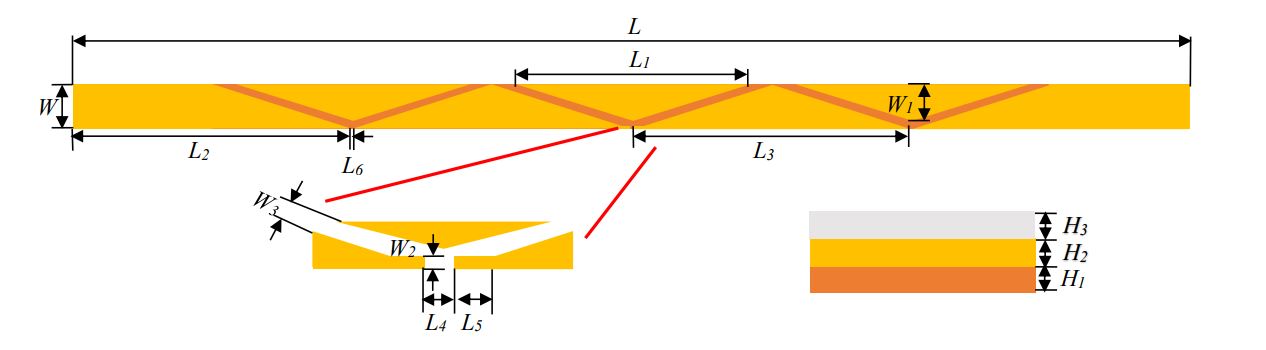}
     \captionsetup{justification=centering}
    \caption{Geometric structure of membrane antenna} 
   \label{fig:geometry}
\end{figure*}

\begin{figure*}
    \centering
    \includegraphics[width=0.9\textwidth]{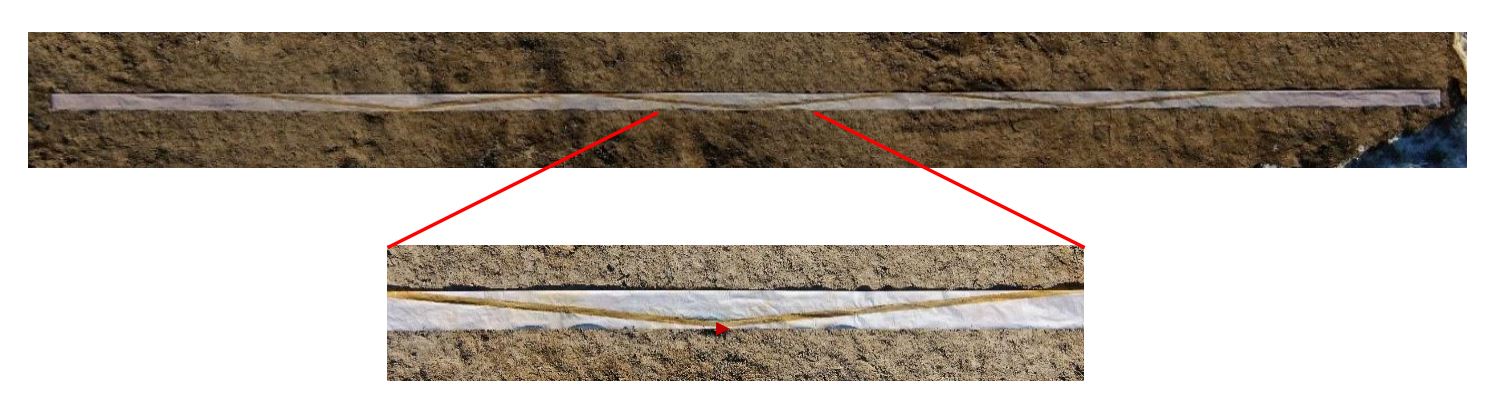}
    \caption{Aerial photograph of a membrane prototype antenna,taken by a drone.} 
    \label{fig:photo}
\end{figure*}

In  the  design  process,  we  first  improve  and  optimize
the antenna structure with various design patterns. We initially  considered  adopting  the  logarithmic  spiral  structure (Figure~\ref{fig:designs}-B),with  a  feed  port  impedance  of  200  $~\Omega$,  capable of achieving ultra-wide band matching. However, for storage  and  deployment  on  the  Moon  by  a  single  rover,  a
long strip of film wrapped around a spool is the most convenient  form to  use.  With  this  in  mind,  we  devote  most of  this  study  to  designs  in  this  form,  such  as  the  dipole square  spiral  and  helical  dipole as shown in Figure\ref{fig:designs}-(C,D), However,  these  still  require  a  large  width  to  work  effectively.

We  find  that  the  planar-coupled  design  can  effectively  expand  the  bandwidth  of  the  antenna.  Planar  coupling  refers  to  certain  conductive  layer  portions  of  the antenna  that  are  not  directly  connected,  but  are  designed around the antenna to change its impedance through interaction  of  the  electromagnetic  field.  The  additional  radiation paths  and  effective  electromagnetic  modes  are  significantly  increased  by  the  planar-coupled  structure, thereby expanding  the  bandwidth  of  the antenna\cite{chen2006broadband}.  By  using a balun  for  impedance  transformation,  a  planar-coupled wide-band dipole membrane antenna can be a good candidate. A possible design is shown in Figure \ref{fig:designs}-E. Two triangular  regions  at  the  bottom  of  the  center  are  radiation blades,  and  other  areas  are  planar-coupled  structure  which are  not  connected  with  each  other.  The  feed  port impedance is 300 $~\Omega$.

To simplify the simulation process and reduce simulation  time,  we  first  consider  the  free  space  background  for the  membrane  antenna  design  and  parameter  optimization. Performance  and  optimization  of  the  membrane  antenna are  then  discussed  and  optimized  with  a  simple  lunar  soil model, based on current lunar measurement data.

\subsection{ Parameter Optimization}
After extensive simulations using the Computer Simulation Technology (CST) and  FEKO(FEldberechnung  beiKörpern  mit  beliebiger  Oberfläche) software suites,  on a number of different designs, we propose a lightweight, foldable,  planar-coupled design. The main body of the membrane antenna is made up of isosceles  triangular-shaped structures  of conducting  area,  with  a  connection balun placed  directly  beneath  the  center.  The  remaining  triangular  structures  are  not  connected  but  they  alter  the  antenna impedance  through  planar  coupling,  thereby  extending  the antenna's  impedance  bandwidth.  This  allows  the  membrane antenna, though a geometrically narrow structure, to have a large impedance bandwidth. 

The sizes of various structures and thickness of the membrane antenna are marked in Figure \ref{fig:geometry} and given in Table \ref{tab:dimension}. The base material use for the bottom layer of the membrane is polyimide. The middle layer is metallic copper acting as the radiating layer, and the top layer is a coating  layer  used  for  preventing  copper  oxidation.  Because this  top  layer  is  very  thin,  it  is  difficult  for  the  software to  simulate  it  accurately.  Because  the  top  layer  of  paint has  minimal  impact  on  the  thin-film  antenna,  it  is  omitted  from  the  simulation.  An  aerial  photograph  of  the actual  prototype,  placed  on  the  ground, is shown in the Figure \ref{fig:photo}.

\begin{table}
\begin{center}
\caption{Dimensions of the membrane antenna}
\label{tab:dimension}
\begin{tabular}{cc|cc}
\hline
    parameters&size(cm)&parameters&size(cm)\\
\hline
     $L$&2000&$W$&25\\
     $L_{1}$&400&$W_{1}$&20\\
     $L_{2}$&499.5&$W_{2}$&1\\
     $L_{3}$&499&$W_{3}$&4.6\\
     $L_{4}$&1&$H_{1}$&1.2*10$^{-3}$\\
     $L_{5}$&1&$H_{2}$&1.2*10$^{-3}$\\
     $L_{6}$&1&$H_{3}$&$<$0.1\\
\hline
\end{tabular}
\end{center}
\end{table}

\begin{figure}[h]
    \centering
    \includegraphics[scale=0.4]{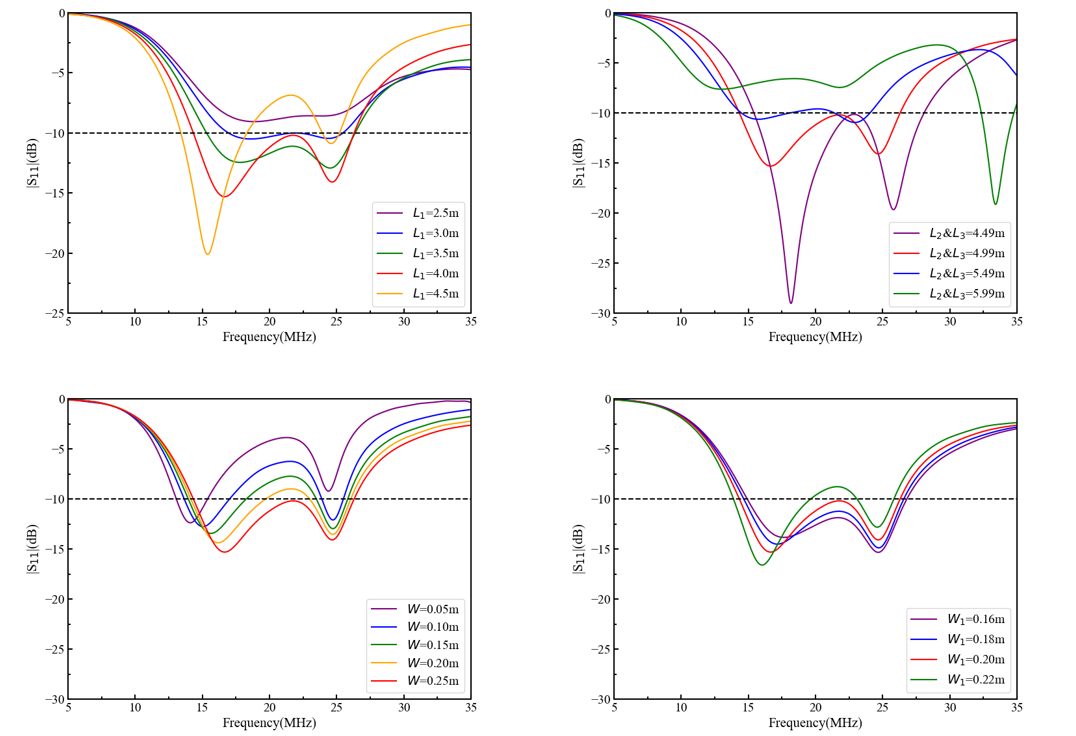}
    \caption{Optimization of membrane antenna parameters.(top left) The parameter scan of L1. (top right) The parameter scans of L2 and L3. (lower left)The parameter scan of W. (lower right)The parameter scan of W1.The red curve represents the optimal parameter set.}
    \label{fig:opt_param}
\end{figure}

The  size  of  the  membrane  antenna  is  determined  by several  parameters.  We  focus  on  scanning  and  optimizing a  few  primary  parameters,  with  the  impedance  bandwidth serving  as  the  principal  criterion,  to  achieve  a  larger
impedance  bandwidth  below  30  MHz,  and  employ  the CST simulation software for parameter scanning. Figure \ref{fig:opt_param} displays the $|S_{11}|$ result of the different $L_{1}$, $W$, $W_{1}$, $L_{2}$ and $L_{3}$.For  this  parameter  set, we find that for the optimal parameter set $L_{1}$=4 m, $W_{1}$=0.2 m,$L_{2}$=4.995 m, and $L_{3}$=4.99 m, the  bandwidth  for  which  $|S_{11}|< -10 \dB$ is  the  widest  ranging  from  15  MHz  to  27  MHz.In addition,  the  simulation  results  show  that  increasing the width of the membrane antenna, $W$, improves $|S_{11}|$, further broadening the bandwidth. However, the width is necessarily  constrained  by  practical  engineering  specifications, and the antenna therefore cannot be overly wide .

\subsection{Balun}

The  membrane  antenna  requires  a  balun  structure  for
balanced-to-unbalanced  conversion,  as  well  as  impedance
transformation. With many types of balun structures available,  choosing  a  suitable  design  is  crucial  to  the  performance  of  the  antenna.  A  transformer  balun  exhibits  superior  performance  in  this  band  and  can  achieve  an  ultralarge  bandwidth,  so  choose  this  variety  of  balun  for  the membrane  antenna  and  use  impedance  transformation  to widen the impedance bandwidth.

\begin{figure}[h]
    \centering
    \includegraphics[width=0.5\textwidth]{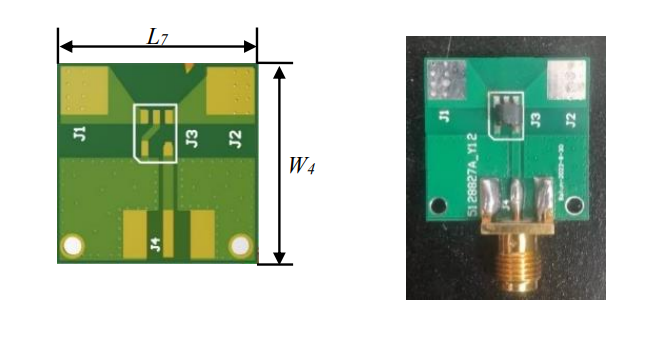}
    \caption{Illustration of the impedance-transforming balun,showing physical dimensions, and a photograph of the balun used.} 
    \label{fig:balun}
\end{figure}

\begin{figure}[h]
    \centering
    \includegraphics[scale=0.4]{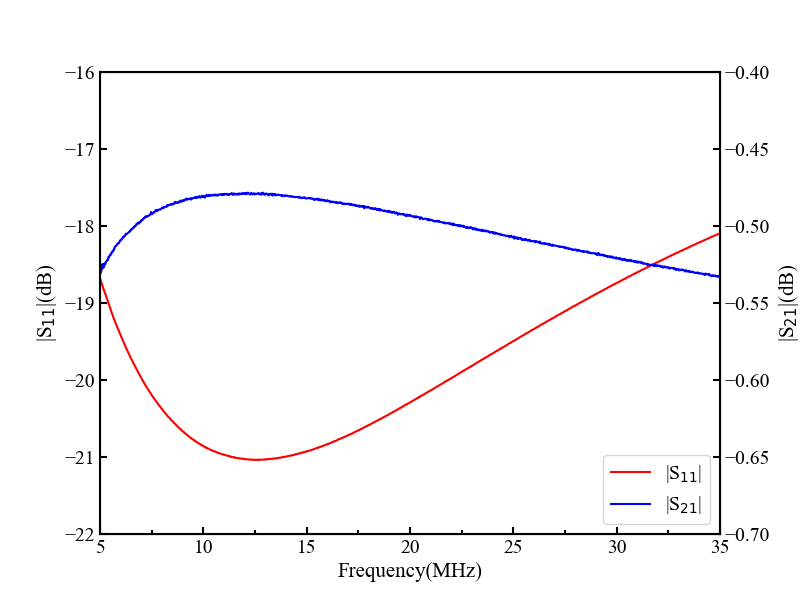}
    \caption{ S-parameter measurement of the balun.The red and
blue lines represent $|S_{11}|$ and$|S_{21}|$ , respectively.} 
    \label{fig:S_balun}
\end{figure}

The  test  model  of  transformer  used  here  is  a  MABA011040,  manufactured  by  MACOM, with  an  impedance transformation  ratio  of  1:6. Figure \ref{fig:balun} shows a photograph of the transformer balun, with dimensions $L_{7}$=2.4 cm and $W_{4}$=3.1 cm. The  size  of  the  balun  was  minimized  to reduce insertion loss and lessen the impact on the mechanical  structure  of  the  membrane  antenna.  Measurement  of
the  balun's  S-parameters  was  performed  using  a  back-to-back  cascade  approach  with  a  Copper  Mountain TR1300/1  vector  network  analyzer  (VNA). Figure \ref{fig:S_balun} shows that $|S_{11}|<-18$ dB, and $|S_{21}| >-0.55$ dB across the entire frequency band. Through actual measurement, the 1:6 transformer balun exhibits excellent impedance matching and low transmission loss, meeting the requirements of the membrane antenna for ultra-wideband.

\section{Simulation}

Here, we consider simulations of our antenna models.We  first  make  a  validation  test  by  simulating  a  membrane  antenna  model  on  the  terrestrial  ground,  which  can be compared  with  the  measurement  data.  We  then describe  the  model  for  the  lunar  soil  and  use  it  to  simulate the performance of the antenna on the lunar surface. 

\subsection{Terrestrial Validation Test }

We first validate our simulation by using terrestrial measurements of the membrane antenna. A membrane antenna with a feed port impedance of $300~\Omega$ is connected via a 1:6 impedance-transforming balun transformer, and tested in an open field near the Hongliuxia observatory in Balikun County, Xinjiang, where the Tianlai experiment is located \cite{Li_2020}. The $|S_{11}|$ parameter is measured with a portable VNA. The relative permittivity and conductivity of the ground at the site is also measured. The measurement of the relative dielectric constant is conducted using the Time Domain Reflectometry method \cite{1999Temperature}, and the conductivity measurement is performed using the Four-point probe method which was originally developed for semi-conductor material \cite{1954Resistivity}. We model the testing site ground as an infinitely large ground plane, with the underground filled by a media with a relative permittivity of 3.60, and a conductivity 0.064 S/m, per our measurements.

\begin{figure}[h]
    \centering
    \includegraphics[scale=0.4]{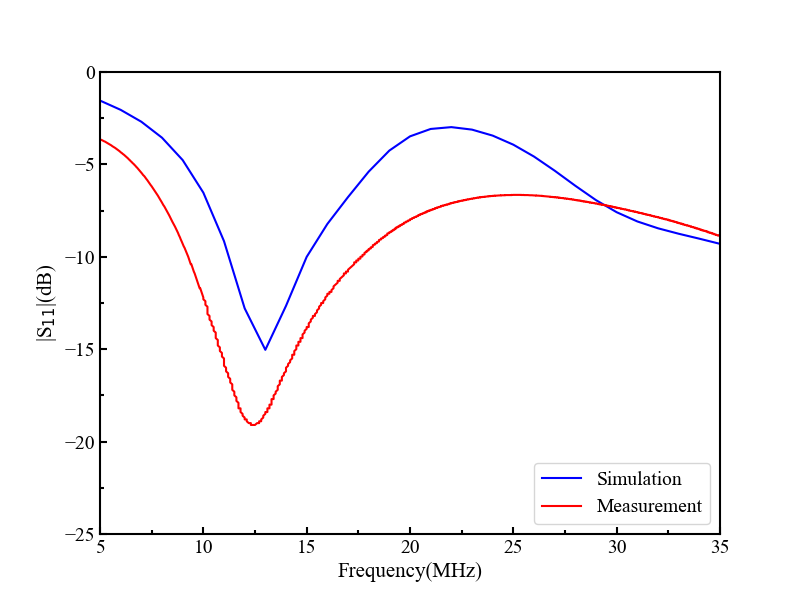}
    \caption{ The measured $|S_{11}|$ on the ground} 
    \label{fig:S11_measurement}
\end{figure}

Figure \ref{fig:S11_measurement} shows the terrestrial measurement and simulation results. showing that the measured $|S_{11}|$ is below -5 dB in the 10-35 MHz frequency band, with a resonance point at around 13 MHz, this is generally consistent with the simulation, though there is still some difference in the magnitude of $|S_{11}|$,  probably due to the simplification  of  the  ground  model.  Real  soil  may  have  different
conductivity  and  permittivity  at  different  depths,  leading
to some discrepancy between the simulation and real measurements. This general agreement between the ground measurement and simulation gives us  confidence that we can use the simulation for our design, provided we use a reasonably good lunar surface model.

\subsection{The Lunar Surface Model}
\label{sec:lunar_surface_model}
To use the membrane antennas on the surface of the moon, it is necessary to understand the impact of the dielectric properties of lunar soil on its performance. 
The dielectric constant is a complex number:

\begin{equation}
    \varepsilon=\varepsilon ' - i\varepsilon ''=\varepsilon_{0}(\varepsilon_{r} -i\frac{\sigma }{\omega\varepsilon_{0}}) 
\end{equation}

where  $\varepsilon {}^{'}$is  the  real  part  of  the  complex  dielectric  constant,  $\varepsilon {}^{''}$is the imaginary part,where $\varepsilon {}_{0}$ is the vacuum permittivity, $\varepsilon {}_{r}$ is the relative permittivity,  $\sigma$ is the electrical conductivity, $\omega$ is the angular frequency. 
The imaginary part of the permittivity is generally characterized using the loss tangent, defined as:
\begin{equation}
   \tan\delta =\frac{\varepsilon''}{\varepsilon '} 
\end{equation}

The dielectric properties of lunar surface materials vary with different locations and depths. from existing research on the dielectric properties of lunar samples obtained from the Apollo and Luna missions\cite{sourcebook1991}, the real part of the permittivity of lunar soil is primarily related to density, showing little correlation with chemical composition and mineral constituents, while the loss tangent is associated with density and the proportion of FeO and TiO{$_2$}\cite{sourcebook1991}:
\begin{equation}
  \varepsilon_{r}=1.919^{\rho } 
\end{equation} 
\begin{equation}  
  \tan\delta ={10^{0.440\rho-2.943}}
\end{equation}
\begin{equation}  
   \quad \tan\delta ={10^{0.038({\rm FeO+TiO_{2}})\%+0.312\rho-3.26}}
\end{equation}

Recent Chinese lunar missions, namely Chang'e 3, 4, and 5, user ground-penetrating radar to infer the complex permittivity of the lunar surface, yielding similar results \cite{Feng2022,Ding2020,Su2022}. Chang'e  4,  in  particular,  was  the  first  to  land on  the  far  side  of  the  moon,  offering  valuable  reference data for future low-frequency exploration. Chang'e 4 measured  the  complex  dielectric  constants  of  the  lunar  surface down to a depth of 10 m, showing the relative permittivity to be between 2.64 and 3.85, with a loss tangent between 0.0032 and 0.0044 \cite{Feng2022}. Chang'e  6  is  intended  to return the first soil samples from the lunar far side, potentially  allowing  the  lunar  surface  model  to  be  further
improved as new data becomes available.

\begin{figure}[h]
    \centering
    \includegraphics[width=0.4\textwidth]{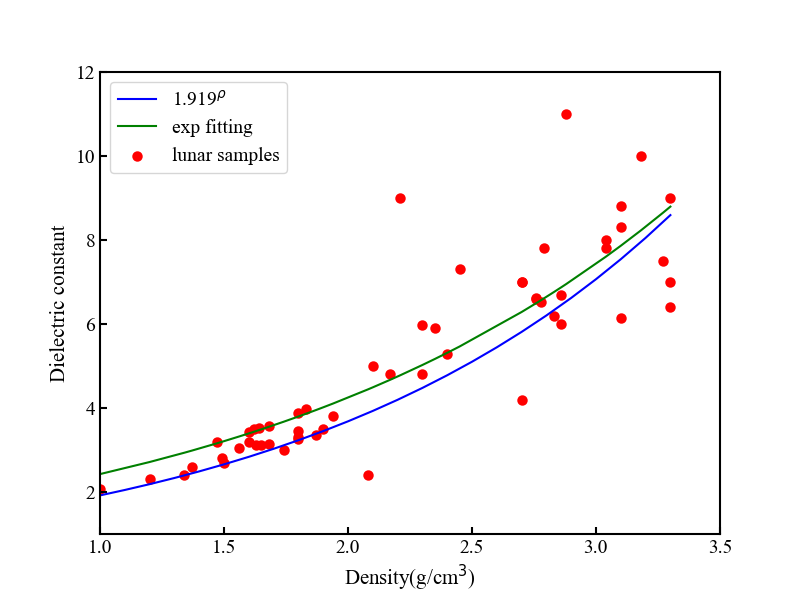}
    \includegraphics[width=0.4\textwidth]{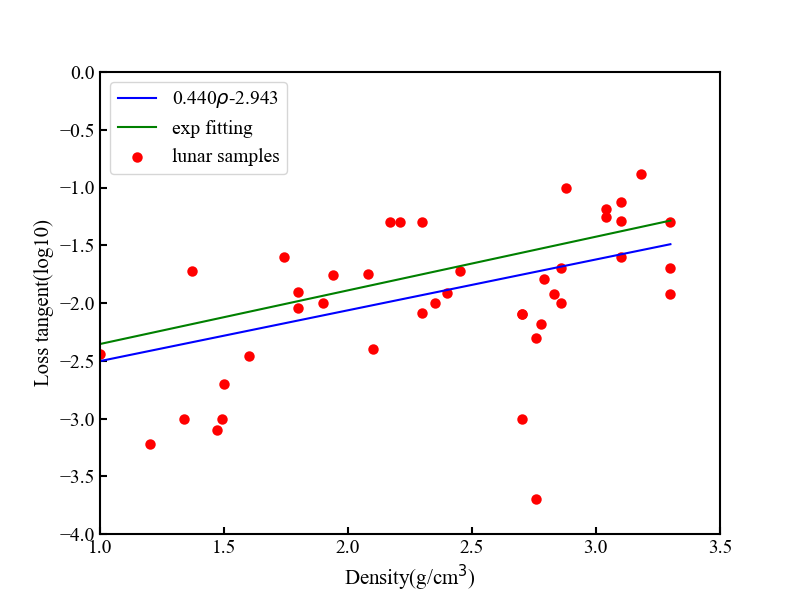}
    \caption{The relative permittivity (top) and loss tangent (bottom) of lunar soil as a function of density } 
    \label{fig:lunar_soil_property}
\end{figure}

\begin{figure}[h]
    \centering
    \includegraphics[scale=0.4]{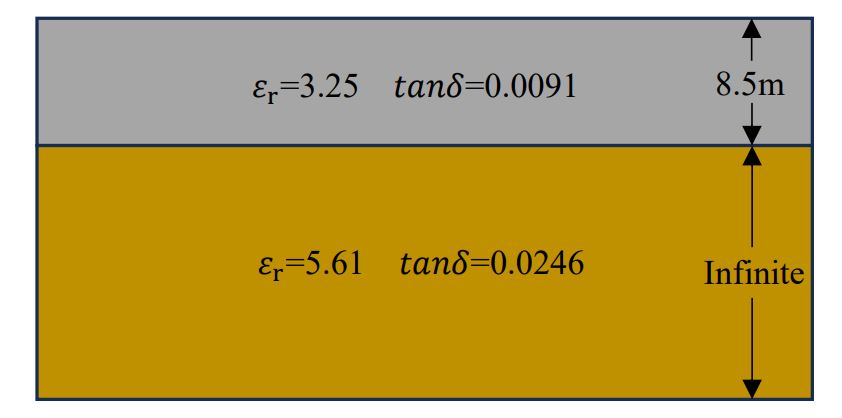}
    \caption{Schematic diagram showing a simplified lunar surface
structure model, with relative permittivities and loss tangents.} 
    \label{fig:lunar_layermodel}
\end{figure}

The permittivity of lunar samples have been measured at frequencies of 0.1 MHz, 1 MHz, and 450 MHz\cite{sourcebook1991}. As shown in Figures \ref{fig:lunar_soil_property}, the measurements of permittivity of lunar samples at 0.1 MHz and 1 MHz are generally in line with the above formula derived from the bulk samples, with the relative permittivity showing strong correlation with density. 

The loss tangent also increases with density; although, the correlation is not as strong, and it is also related to the material content, particularly ilmenite concentration \cite{Zhu2021}. The loss tangent also changes with frequency, with the lowest resonant point occurring around 10 MHz, which is also temperature-dependent\cite{sourcebook1991}. On the whole, the loss tangent on the lunar surface is extremely small, indicating excellent insulation and for membrane antennas their radiation efficiency would not be much affected.

The lunar surface material consists of both regolith and rocks. Here, we model it as two plane layers,  using the FEKO infinite multilayered medium simulation. The first layer consists of weathered regolith , which typically has an average thickness of 4-5m in lunar maria and 10-15m in highland regions\cite{Mckay1991}. Here we consider the average  thickness  of  8.5 m,  with  density  increasing  gradually from  1.49  at  the  surface  to  2.07 g·cm$^{-3}$ at a depth of 8.5 m.  With depth,the corresponding relative permittivity increases from 2.64 to 3.85 \cite{Feng2022}, with an average value of 3.25. This aligns with the sample's average loss tangent of 0.0091. The second layer, the rock layer, is modeled with an effectively infinite thickness. Its complex permittivity is primarily derived from Chang'e 4's ground-penetrating radar measurements, with a density in the range of 2–2.6  g·cm$^{-3}$at depths of 10–45 m.  The average relative permittivity is taken to be 5.61,based  on  the  data  presented in Figure \ref{fig:lunar_soil_property}, with a loss tangent of 0.0246 and thickness  taken  as  infinite.  This  structure  is  illustrated  in Figure \ref{fig:lunar_layermodel}.


\begin{figure}[h]
    \centering
    \includegraphics[scale=0.37]{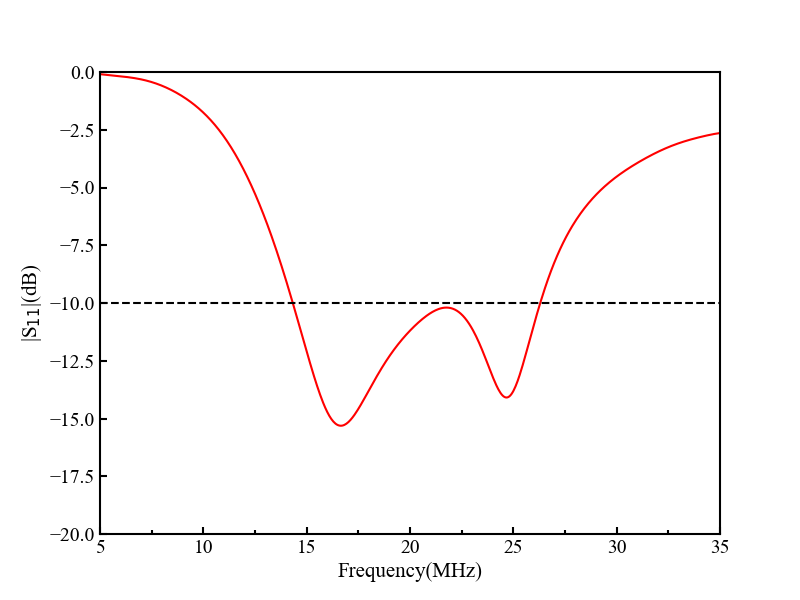}
    \includegraphics[scale=0.37]{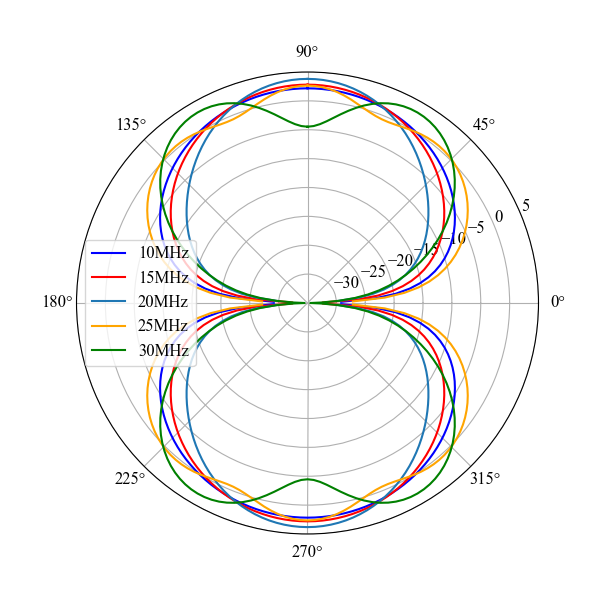}
    \includegraphics[scale=0.37]{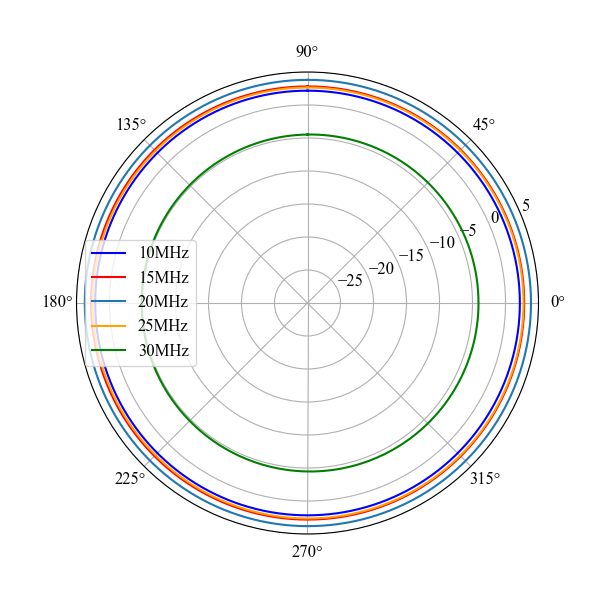}
    \caption{Simulation results for a membrane antenna in free space. Top Panel: the $|S_{11}|$ parameter, Middle Panel: the power beam pattern in E-plane, Bottom Panel: the power beam pattern in H-plane. }
    \label{fig:pattern_freespace}
\end{figure}

\begin{figure}[t]
    \centering
    \includegraphics[width=0.35\textwidth]{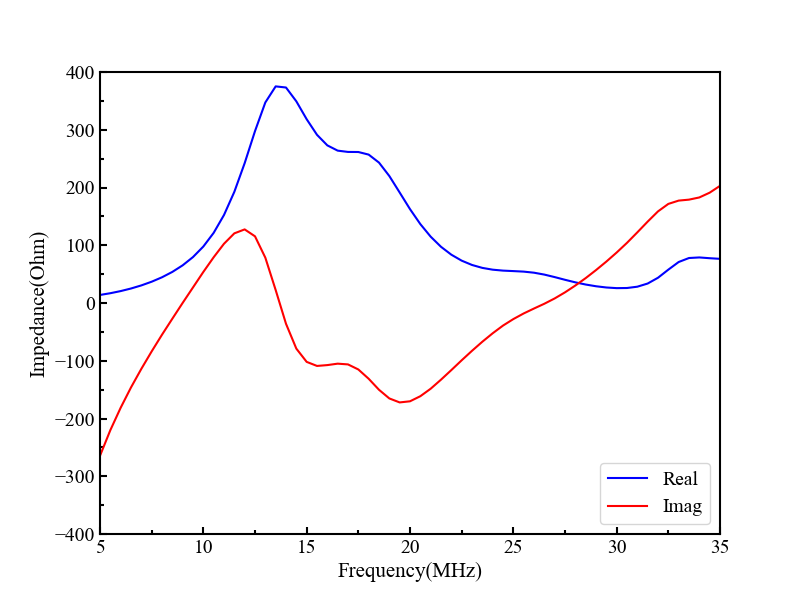}
    \includegraphics[width=0.35\textwidth]{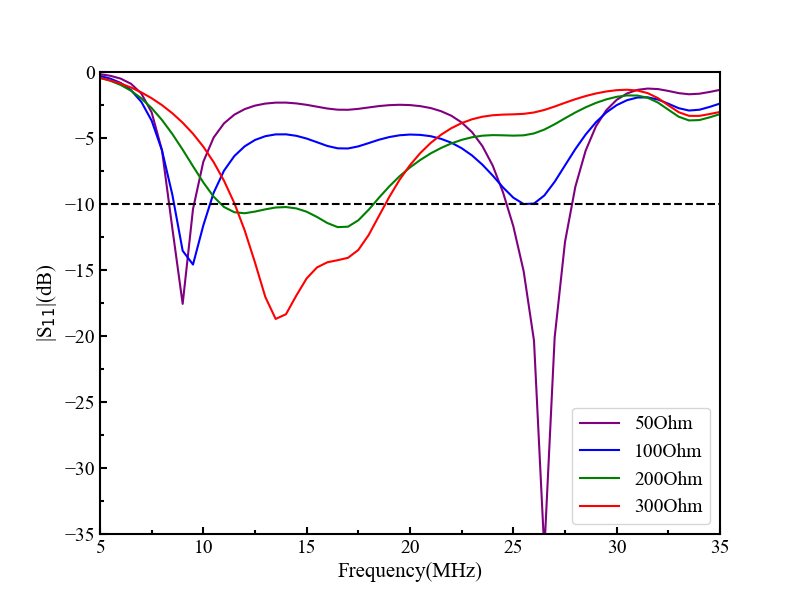}
    \caption{Top:Antenna impedance on lunar surface, Bottom: $|S_{11}|$ for various reference impedance values.} 
    \label{fig:ant_imp}
\end{figure}

\begin{figure}[h]
    \centering
    \includegraphics[scale=0.35]{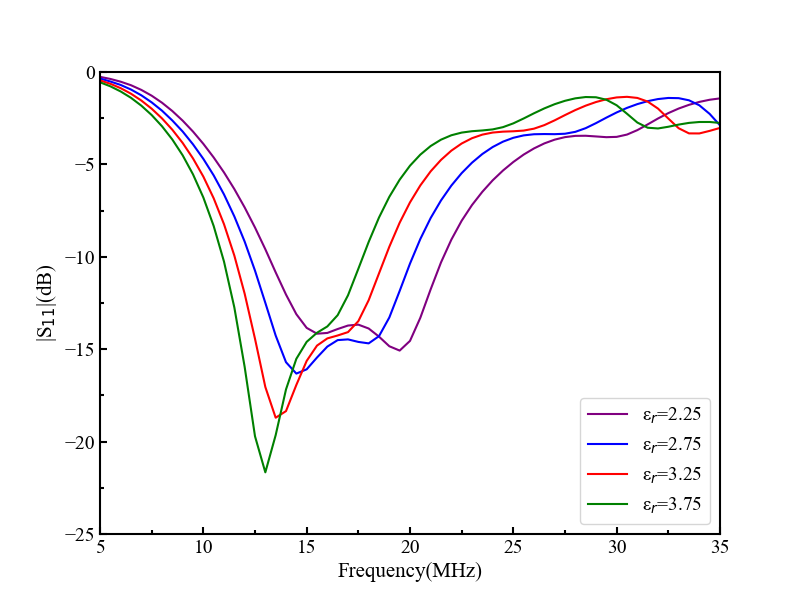}
    \includegraphics[scale=0.35]{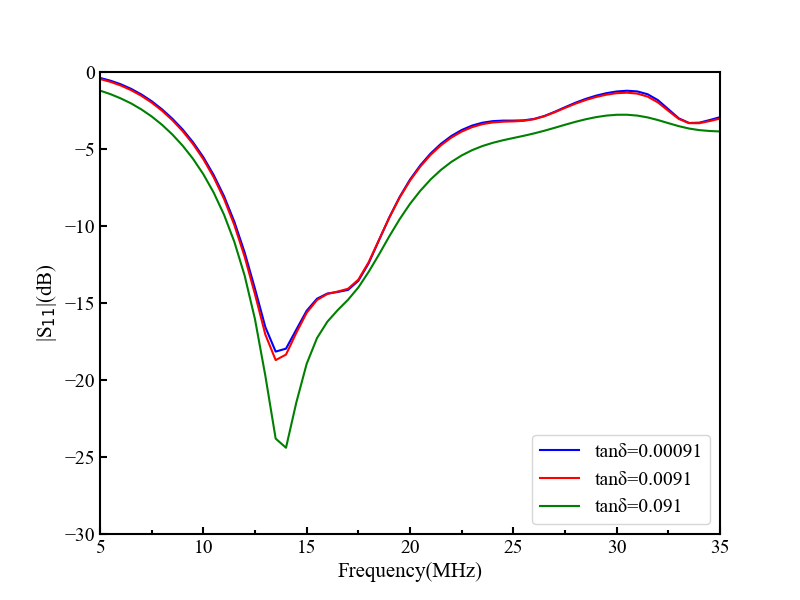}
    \caption{Comparison of $|S_{11}|$ for different relative permittivity(top) and loss tangent (bottom).} 
    \label{fig:var_perm_tan}
\end{figure}

\subsection{Simulation with the Modeled Lunar surface}

For a simulation in free space, the $|S_{11}|$ parameter and the power beam pattern are shown in Figure \ref{fig:pattern_freespace}.yielding $|S_{11}|$ is below -10 dB in the 15-27 MHz frequency band with a relative bandwidth of 57\%.  As the membrane antenna designed in this work is a type of dipole antenna, the directional pattern is essentially consistent with that of a dipole, maintaining a characteristic a wide beam  throughout the entire frequency band with a peak gain of about 2.5 dBi. The E-plane directional pattern splits slightly at the maximum frequency of 30 MHz. This occurs because the antenna size exceeds the half wavelength, resulting in the splitting of the main lobe due to the presence of reverse currents, though the direction of the lobes does not change significantly. The H-plane directional pattern is essentially omni-directional in the entire frequency band. 

Our simulation uses the simplified lunar surface model described above, in Section \ref{sec:lunar_surface_model}. The top panel of Figure \ref{fig:ant_imp} shows the simulated antenna impedance, including both the real and imaginary parts. For most of the frequency range, the imaginary part is negative,giving a mostly capacitive impedance. The bottom panel of \ref{fig:ant_imp} 
shows the $|S_{11}|$ under different reference impedances. The widest impedance-matching bandwidth with $|S_{11}|<-10 \dB$ is achieved when the impedance transformation ratio  is 1:4 on the lunar surface. $|S_{11}|$ is below -10 dB within the frequency range of 11-18 MHz, yielding a relative bandwidth of 48.2\%. However, as mentioned before, the maximum impedance bandwidth occurs under a 1:6 impedance transformation in free space, indicating that the optimal balun impedance transformation ratio varies between free space and the lunar surface.


In the simplified lunar surface model, shown in Figure \ref{fig:lunar_layermodel}, the dielectric properties of the first layer of lunar soil (or regolith) predominantly affect the performance of the membrane antenna. However, because the dielectric properties of the lunar surface vary with location and depth. We examine a range of relative permittivity and loss tangent. Figure \ref{fig:var_perm_tan} shows the influence of complex dielectric properties on the performance of the membrane antenna on lunar surface. The relative permittivity affects the $|S_{11}|$ parameter of the antenna significantly. As it increases, the resonance points shift towards lower frequencies,as expected. The influence of loss tangent on $|S_{11}|$ is less noticeable, mainly due to the very low loss tangent value in the regolith. Observable changes only occur when they reach to a threshold level.

\begin{figure*}
    \centering
    \includegraphics[scale=0.4]{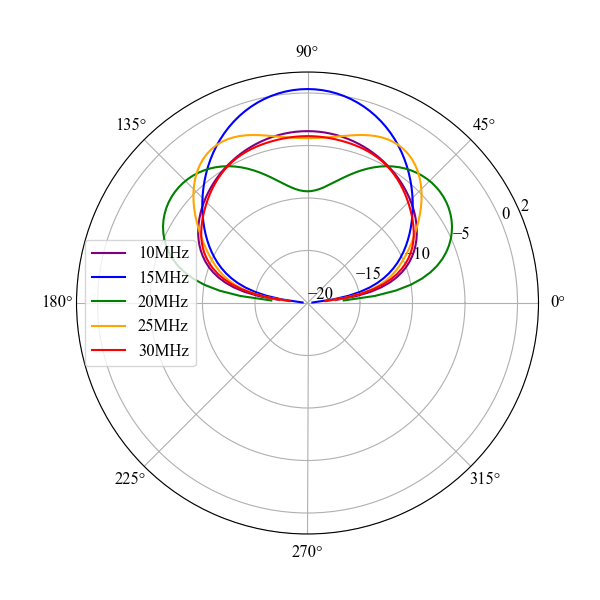}
    \includegraphics[scale=0.4]{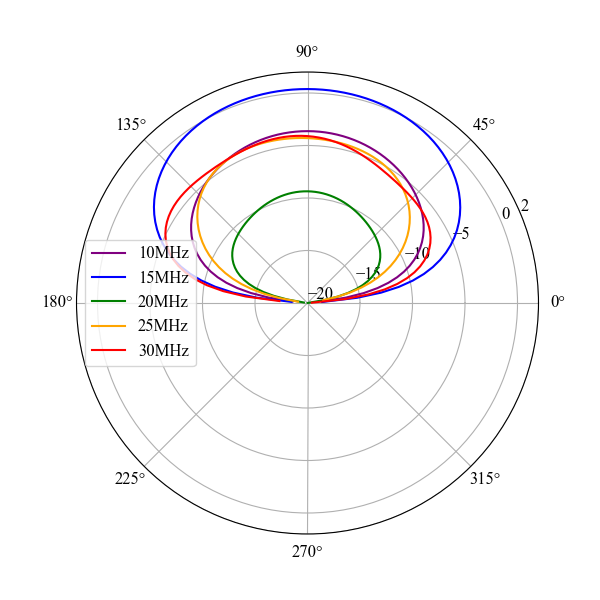}
    \caption{The beam pattern of the membrane antenna model. Left: E-plane, Right: H-plane. } 
    \label{fig:beampattern}
\end{figure*}

Simulated beam patterns are shown in Figure \ref{fig:beampattern}. These beam peaks in the zenith direction, with a peak gain is less than 1 dBi, somewhat smaller than the free space case due to Ohmic loss on the lunar surface. There are some ripples modulating on top of the dipole pattern due to apparent chromaticity, i.e. the beam pattern changes with frequency. These effects are relatively minor and can be easily accommodated in conventional astronomical observations. However, observation of 21 cm fluctuations in the cosmic dark ages requires extremely high precision, and such variations and ripples may introduce complications. Further efforts will be required to compensate these effects.   


\begin{figure*}
    \centering
    \includegraphics[scale=0.4]{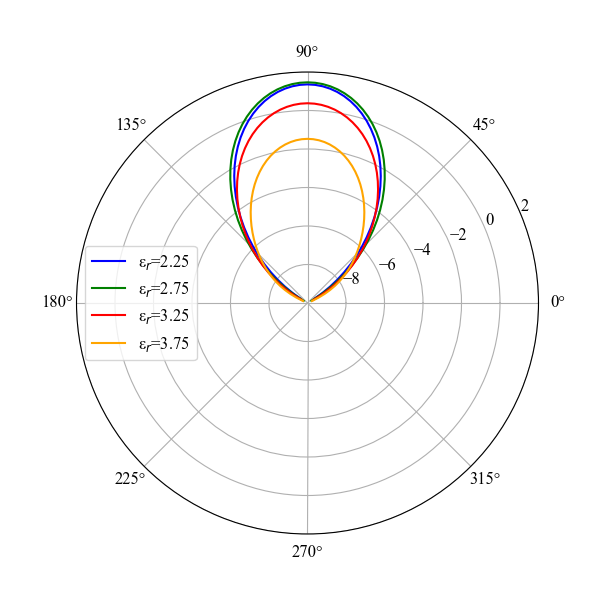}
    \includegraphics[scale=0.4]{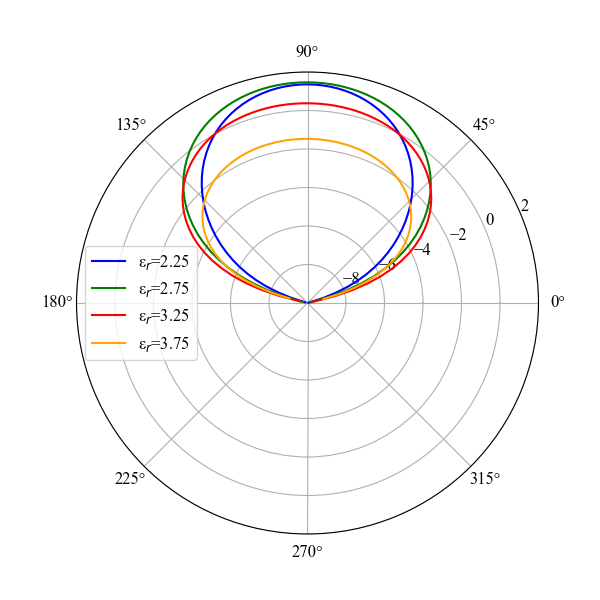}
    \caption{Effects of relative permittivity on beam pattern. Left: E-plane, Right: H-plane. } 
    \label{fig:beampattern_perm}
\end{figure*}
\begin{figure*}
    \centering
    \includegraphics[scale=0.4]{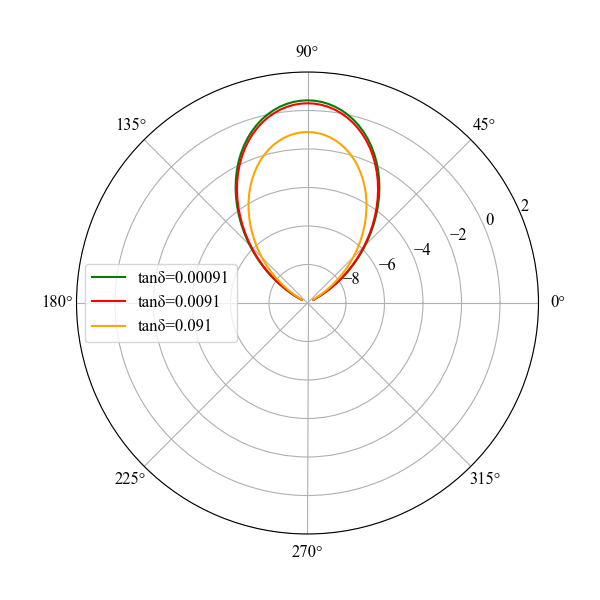}
     \includegraphics[scale=0.4]{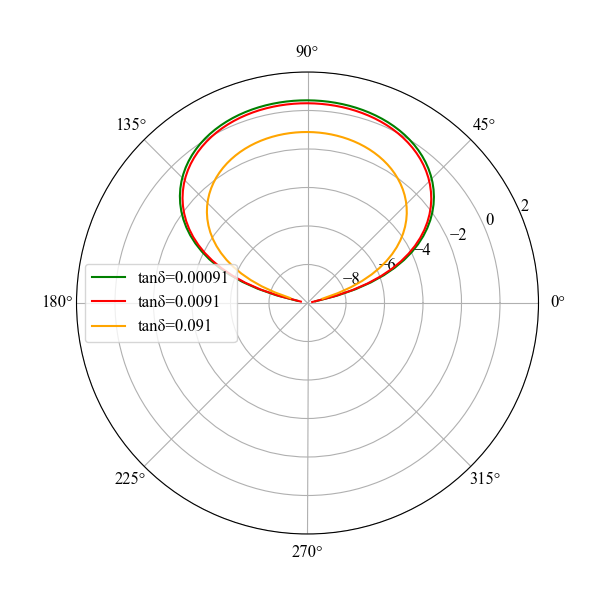}
    \caption{Effects of the loss tangent on beam pattern. Left: E-plane, Right: H-plane.} 
    \label{fig:loss_tan}
\end{figure*}


Figure \ref{fig:beampattern_perm} shows the impact of different permittivity values on the beam pattern at 15 MHz. An increase in relative permittivity results in increased ground loss and decreased gain for the membrane antenna in both the E-plane and the H-plane. 

the pattern also changes with the variation of the loss tangent, as shown in Figure \ref{fig:loss_tan},with an increase in loss tangent leading to a weak decrease in antenna gain. This influence is negligible while the loss tangent is below a threshold level, becomes very obvious at higher values.


\subsection{Antenna efficiency}
Antenna efficiency here is determined primarily by antenna Ohmic loss, ground loss and impedance mismatching. For membrane antenna on lunar surface, Ohmic loss and ground loss are both small, making impedance mismatching the dominant consideration for antenna efficiency.  Efficiency from impedance mismatching is calculated using the relationship
\begin{equation}
   \eta=1-|\Gamma^2|
\end{equation}
where $\eta$ is  the  antenna  efficiency  and $\Gamma$ is  the  reflection
coefficient.

For frequency bands below 35 MHz, the sky signal is mostly from the synchrotron radiation of the Milky Way. The brightness temperature of the synchrotron radiation exceeds $ \sim 10^4$ K at 30 MHz, and can be in excess of $\sim 10^7$ K at a few MHz, so it is the dominant component of system temperature. For an antenna efficiency of 10\%, the signal-induced antenna temperature still far exceeds the environment temperature of the system, so a good signal-to-noise ratio can still be obtained. we take the bandwidth of the 10\% antenna efficiency as workable bandwidth for the lunar antenna.

\begin{figure}[t]
    \centering
    \includegraphics[scale=0.4]{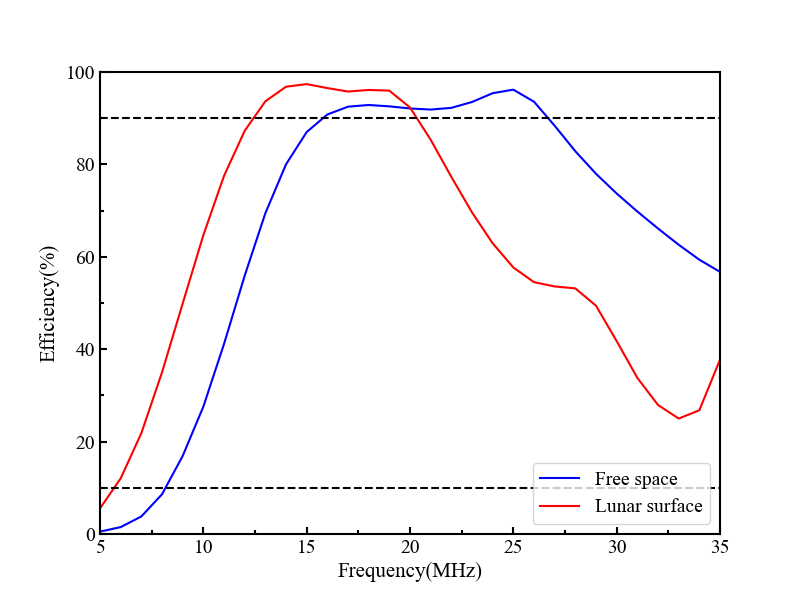}
    \caption{Comparison of antenna efficiency of membrane antennas in free space and on the lunar surface } 
    \label{fig:ant_eff}
\end{figure}

Figure \ref{fig:ant_eff} shows a comparison between the efficiency in free space and on the lunar surface, indicating that this membrane antenna can achieve efficiency greater than 10\% in free space at 7.4-35 MHz and  efficiency greater than 90\% at 15-27 MHz. The lunar surface simulation results show that the membrane antenna's efficiency exceeded 10\% at 5-35 MHz and the efficiency exceeded 90\% at 12-19 MHz.

\section{Summary}
 Lunar-based radio astronomy, as an important astronomical  goal,  is crucial for studying the epochs of the Cosmic Dark Ages and Cosmic  Dawn.  In such projects, the membrane antenna could play an important  role.  It has a number of favorable properties, such as light weight, compact collection volume,  easy deployment, and no need to be  fastened against wind in the lunar environment, making it well suited  for use on a lunar mission. The receiving circuit can be directly  mounted on the membrane antenna, forming an integrated receiver.

In  this  study,  we  investigate  the  electrical  design  of
membrane  antennas,  noting  that  polyimide  films  have
been  used  for  previous  space-based  applications.  However, in using this material for an antenna on the lunar surface,  performance  and  durability  still  need  to  be  investigated under the extreme variations in temperature and radiation  exposure  present  on  the  surface  of  the  moon.  The goal  is  to  construct  a  large  radio  interferometer  array  for ultra-long-wavelength  astronomy,  i.e.,  at  frequencies below  30  MHz.  The  antenna  unit  must  have  a  wide  beam and  a  broad  working  bandwidth,  in  order  for  the  array  to
achieve a large field of view with the potential to be used
in a wide variety of astronomical research, including observing  the  early  universe.  Here,  we  primarily  consider
designs  in  the  form  of  a  long,  narrow  strip,  as  the  most
practical form to store, transport, and deploy.

Our  antenna  adopts  a  planar-coupling  broadband
dipole  design,  capable  of  achieving  an  antenna  efficiency
greater  than  10\%  in  free  space  from  7.4–35  MHz  and
greater  than  90\%  antenna  efficiency  from  15–27  MHz
while  maintaining  a  very  wide  radiation  pattern  in  frequency bands below 30 MHz. Informed by results from previous  lunar  missions, we  create  a  simplified  lunar  soil model  and  simulate  the  membrane  antenna  performance when  successfully  deployed.  On  the  lunar  surface,  the antenna  efficiency  is  predicted  to  be  greater  than  10\%  at 5–35  MHz  and  greater  than  90\%  at  12–19  MHz,  with  the radiation  pattern  maintaining  a  dipole  form  at  5–30  MHz.As  noted  in  Section  3.3,  for  observations  at  such  low  frequencies, the sky temperature is very high, meaning that a
10\% antenna efficiency is sufficient for astronomical observations with a good signal-to-noise ratio.

This study still has a number of limitations.  It assumes the antenna membrane to be laid on a perfectly flat and uniform ground plane and adopts a highly simplified model for  the lunar ground.  In reality,  it is unlikely to be deployed on a flat surface, and a more complex distribution  of  subsurface  material  may  be  present.  While  we
vary  our  ground  model  parameters,  such  more  complicated cases are not considered here. Additionally, the membrane  antenna  itself  may  also  have  bends  and  folds.  Such issues  will  be  addressed  in  future  work.  Nevertheless,  the work  presented  here  provides  a  good  foundation  for  the design  of  a  lunar-based  radio  telescope  and  can  be  used to  aid  in  the  design  of  antenna  arrays  on  the  lunar  far side.

\section*{acknowledgements}

We acknowledge the support of the National SKA program of China(Nos. 2022SKA0110100 and 2022SKA0110101), the Natural Science Foundation of China(Nos. 12273070,12203061, 1236114814, 12303004).

\section*{Author Contributions}
Xuelei Chen conceived the basic idea of the project, Fengquan Wu guided Suonanben to conduct the simulation and experimental work, Tianlu Chen participated in the discussion. Kai He, Shijie Sun, Wei Zhou,  Minquan Zhou, Cong Zhang, Jiaqin Xu, Qisen Yan, Shenzhe Xu, Jiacong Zhu, Zhao Wang,  Ke Zhang and  Wang Yougang made contributions in field testing. Li Jixia provided support in data acquisition and storage.   Miao Haitao offered assistance in programming and paper revision. All authors read and approved the final manuscript.
 
\section*{Declaration of Interests}
The authors declare no competing interests.
Please check the Instructions for Authors of the journal to see the contents must be contained here.

\appendix                  



\bibliographystyle{unsrt} 
\bibliography{ati}      

\label{lastpage}
\end{document}